\documentclass{article}
\usepackage[dvips]{graphicx}
\usepackage{graphics}
\usepackage{amssymb}
\usepackage{bm}
\usepackage{amsmath}
\usepackage{hyperref}
\def\spose#1{\hbox to 0pt{#1\hss}}

\def\lta{\mathrel{\spose{\lower 3pt\hbox{$\mathchar"218$}}
     \raise 2.0pt\hbox{$\mathchar"13C$}}}
\def\gta{\mathrel{\spose{\lower 3pt\hbox{$\mathchar"218$}}
     \raise 2.0pt\hbox{$\mathchar"13E$}}}
\newcommand{\be}{\begin{equation}}
\newcommand{\en}{\end{equation}}
\newcommand{\bea}{\begin{eqnarray}}
\newcommand{\ena}{\end{eqnarray}}

\begin{document}

\title{Bohmian Quantization of the Big Rip}
\author{Nelson Pinto-Neto\footnote{nelsonpn@cbpf.br} and Diego Moraes Pantoja\footnote{diegomp@cbpf.br}\\
\textsl{\small{ICRA - Centro Brasileiro de Pesquisas F\'{\i}sicas -- CBPF,}}\\
\textsl{\small{rua Xavier Sigaud, 150, Urca, CEP22290-180, Rio de Janeiro, Brazil}}}

\date{\today}

\maketitle

\begin{abstract}

It is shown in this paper that minisuperspace quantization of homogeneous and isotropic geometries
with phantom scalar fields, when examined in the light of the Bohm-de Broglie interpretation of quantum mechanics,
does not eliminate, in general, the classical big rip singularity present in the classical model. For some values of the Hamilton-Jacobi 
separation constant present in a class of quantum state solutions of the Wheeler-DeWitt equation, 
the big rip can be either completely eliminated or may still constitute a future attractor for all expanding solutions.
This is contrary to the conclusion presented in Ref.\cite{kie}, using a different interpretation of the wave function, 
where the big rip singularity is completely eliminated ("smoothed out") through quantization, independently 
of such separation constant and for all members of the above mentioned class of solutions. 
This is an example of the very peculiar situation where different interpretations of 
the same quantum state of a system are predicting different physical facts, instead of just giving different 
descriptions of the same observable facts: in fact, there is nothing more observable than the fate of the whole Universe.


\end{abstract}

\maketitle

\section{Introduction}

The discovery that the Universe is experiencing a phase of accelerated expansion \cite{supernova}
constitutes a big challenge to theoretical physics and Cosmology. The simplest explanation,
the existence of a positive cosmological constant, has problems with quantum field theory \cite{cc}.
Other possibilities have been investigated, either by proposing new material sources, called generically dark energy 
\cite{quint},
or through the modification of the gravitational interaction itself \cite{fder}. Even
the whole framework of the standard cosmological model has been questioned \cite{inh}. It is 
expected that these possibilities will be tested by future observations \cite{des}.

One bonus of this revolution was the definite fall, initiated with the advent of inflationary theory, 
of ancient theoretical prejudices based on the postulation
of energy conditions. Nowadays we know many plausible frameworks
where such energy conditions can be violated. Even the possibility of existence of phantom fields,
violating the less severe of these conditions, the null energy condition, has been investigated and some scenarios
proposed \cite{phantom}. The presence of such phantom fields may impose the
existence of a future singularity, called big rip, a rupture of spacetime due to an infinite accelerated expansion,
which can destroy the Universe in some tens of billion years \cite{bigrip}.

One natural question, which has already been asked, concerns the importance of quantum effects as one approachs the big rip:
in the same way that the initial big-bang singularity may be avoided through quantum effects \cite{qc}, one may expect
that the big rip can be circumvanted due to quantum gravitational effects. This possibility has been investigated
in the framework of phantom hydrodynamical fluis \cite{niv} and phantom scalar fields \cite{kie}. In the first,
it is argued that the big rip is not eliminated, while in the second
it is suggested that the big rip is indeed avoided through quantum effects.

In order to arrive at these conclusions, one must extract the relevant information from the wave function of the 
Universe according to a definite interpretation of quantum mechanics.
However, as it is well known, when one tries to quantize the whole Universe, one cannot rely on the Copenhaguen interpretation
of quantum mechanics because it imposes the existence of a classical domain outside 
the quantized system in order to generate the
physical facts out of the quantum potentialities. Of course, if we want to quantize the whole Universe, 
there is no place for a classical domain outside it, and the Copenhaguen
interpretation cannot be applied. Some alternatives are the
Bohm-de Broglie \cite{bohm} and the many worlds interpretation 
of quantum mechanics \cite{everett}, where no classical domain is necessary
to generate the physical facts out of potentialities. Hence, they can be applied to the Universe as a
whole. 

In Ref.\cite{kie}, it is argued that the wave packet solution of the Wheeler-De Witt equation
corresponding to the quantized phantom scalar field model, which is peaked around the classical
solution, spreads as one approaches the big rip region of configuration space, indicating
the breakdown of the Born-Oppenheimer approximation and the absence of a WKB time. 
Hence, the semiclassical approximation together with the notion of classical trajectories are not valid any more, and
the classical singularity theorems do not apply: the big rip singularity is "smoothed out" due to quantum effects.

There are two reasons which make this reasoning not conclusive. First of all, spreading of a gaussian
does not necessarily mean that one is out of the classical region. For instance, a gaussian peaked around 
the classical trajectory $x-pt/m$ describing a free quantum particle with momentum $p$ and mass $m$
spreads due to the increasing  of the uncertainty on its position as time passes, approaching the classical
statistical distribution of an ensemble of free particles with classical gaussian distribution of initial
positions. In fact, there are many classical statistical distributions which spreads in time, hence 
spreading of distributions is not necessarily connected with quantum behaviour (see Ref.\cite{hol}, chapter 6,
for a discussion on this subject).
Secondly, even assuming that a classical singularity has really been "smoothed out" by quantum effects,
there is still the possibility of existence of some sort of singularity in the region of configuration space
where the semiclassical approximation is not valid, even though it is rather obscure what should be a definition 
of a quantum spacetime singularity in this picture.

These ambiguities are not present if one uses the Bohm-de Broglie interpretation of quantum cosmology. 
Here, the classical domain is simply defined as the region in configuration space where the Bohmian
trajectories approach the classical ones, which happens when the wave function phase approaches the
classical action or, equivalently, when the so called quantum potential becomes negligible. Also, as
in this interpretation one has a definite notion of trajectory in the classical and quantum domains, 
the singulatity may be defined straightforwardly, even at the quantum level. For instance, in a homogeneous
and isotropic minisuperspace model, a Bohmian scale factor trajectory can be calculated, and
the big-bang singularity is defined as the finite proper time moment where the Bohmian scale factor goes
to zero. An illustrative example is the model described in Ref.\cite{tovar}, where a singularity appears
when quantum effects are present.

The aim of this paper is to reexamine the investigation performed in Ref.\cite{kie} within 
the Bohm-de Broglie interpretation of quantum cosmology. We will show that, using this interpretation, the big rip singularity
is not avoided in general: for a large class of wave solutions of the Wheeler-DeWitt equation it is still
present, even though the wave function spreads in all cases.

The paper is divided as follows: in the next section we describe the classical model where the big rip
appears. In Section 3 we quantize the system and interpret the solutions using the Bohm-de Broglie interpretation 
of quantum cosmology, showing that within this interpretation the big rip is not always avoided.
We end in section 4 with a discussion of the results and their significance, specially when compared
with the results presented in Ref.\cite{kie}, and a conclusion.

\section{The classical phantom cosmology}

Let us take the Einstein-Hilbert action minimally coupled to a scalar field $\phi$, 
\begin{equation}
\label{acao}
	S=-\frac{3}{\kappa ^{2}}\left(\int_{M}d^{4}x\sqrt{-g}R+2\int_{\partial M}d^{3}xh^{\frac{1}{2}} K \right)+S_{\phi},
\end{equation}
where, $k^{2}=8\pi G$, $c=1$, $R$ is the curvature scalar, $h$ is the determinant of the space metric, $K$ is the trace of the extrinsic curvature $K_{ij}$ at the boundary $\partial M$ of the four-dimensional manifold \textit{M}. The
scalar field is accelerating the Universe \cite{supernova}, and it is phantomic ($p/\rho<-1$) \cite{phantom}. The scalar field action
$S_{\phi}$ reads
\begin{equation}
\label{scf}
S_{\phi}=\int_{M}d^{4}x\sqrt{-g}\left[-\frac{1}{2}g^{\mu\nu}\partial_{\mu}\phi\partial_{\nu}\phi-V\left(\phi\right)\right],
\end{equation}
where $g$ is the determinant of the spacetime metric $g_{\mu\nu}$. Note the wrong sign in the kinetic term, causing the phantomic 
acceleration. The potential is given by \cite{exp}
\begin{equation}
\label{potencial}
V=V_{0}e^{-\lambda k\phi}.
\end{equation}

We will work with a spatially flat minisuperspace Friedmann model with homogeneous and isotropic phantom field. Actions (\ref{acao}) 
and (\ref{scf})then read
\begin{equation}
\label{lagragean}
S=-\int dt \left(\frac{3}{N\kappa ^{2}}\dot{a}^{2}a+\frac{1}{2N}a^{3}\dot{\phi}^{2}+Na^{3}V_{0}e^{-\lambda k\phi}\right),
\end{equation}
where $N$ is the lapse function and $a$ is the scale factor.

The canonical momenta are
\begin{eqnarray}
\label{momentoa}
\pi_{a}&=&-\frac{6}{N\kappa ^{2}}\dot{a}a,\nonumber \\
\pi_{\phi}&=&-\frac{a^{3}\dot{\phi}}{N},
\label{momentophi}
\end{eqnarray}
yielding the canonical hamiltonian 
\begin{equation}
\label{hamiltonian}
{\rm H}=N{\cal{H}}=N\left(-\frac{\kappa ^{2}}{12a}\pi^{2}_{a}-\frac{1}{2a^{3}}\pi^{2}_{\phi}+a^{3}V_{0}e^{-\lambda \kappa\phi}\right),
\end{equation}
which is constrained to be null. The hamiltoian constraint ${\cal{H}}=0$ reduces, when written in terms of
the velocities and taking $N=1$, to the Friedmann equation 
\begin{equation}
\label{hubble}
	\left({\frac{\dot{a}}{a}}\right)^{2}\equiv H^{2}=\frac{\kappa^{2}}{3}\left(-\frac{\dot{\phi}^{2}}{2}+V_{0}e^{-\lambda \kappa\phi}\right).
\end{equation}
The energy density and pressure associated with the phantom field are
\begin{equation}
	\rho=-\frac{1}{2}\dot{\phi}^{2}+V\left(\phi\right),
\end{equation}
\begin{equation}
		p=-\frac{1}{2}\dot{\phi}^{2}-V\left(\phi\right).
\end{equation}
The Euler-Lagrange equations read, after combination with the constraint (\ref{hubble}),
\begin{eqnarray}
\label{eqfriedmann}
{\frac{\ddot{a}}{a}}-\frac{\kappa ^{2}}{3}\left(\dot{\phi}^{2}+V_{0}e^{-\lambda \kappa\phi}\right)&=&0,\nonumber \\
\ddot{\phi}+3\frac{\dot{a}}{a}\dot{\phi}+V_{0}\lambda \kappa e^{-\lambda \kappa\phi}&=&0.
\label{pphi}
\end{eqnarray}
An attractor solution can be found \cite{attr} given by
\begin{eqnarray}
\label{phi1}
\phi(t)&=&\frac{2}{\lambda \kappa}\ln\left[1-\frac{\lambda^{2}}{2}H_{0}\left(t-t_{0}\right)\right],\nonumber \\
\alpha(t)&=&-\frac{2}{\lambda^{2}}\ln\left[1-\frac{\lambda^{2}}{2}H_{0}\left(t-t_{0}\right)\right] + \alpha_0,
\label{alpha1}
\end{eqnarray}
where $\alpha(t)\equiv\ln a(t)$. 
This is a big rip solution because at the finite time $t\rightarrow t_{Rip}=t_{0}+\frac{2}{\lambda^{2}H_{0}}$,
the scale factor, the Hubble parameter, and the energy density 
\begin{equation}
\label{rho}
\rho=\rho_{0}\left(\frac{a}{a_{0}}\right)^{\lambda^{2}},
\end{equation}
diverges, and consequently the curvature of spacetime. 

As we will exhibit numerical calculations, from now on we will take $\lambda =2/\sqrt{6}$, $V_0=1/2$, $\kappa =\sqrt{6}$,
which implies $H_0=3/\sqrt{10}$ and, from Eq.(\ref{phi1}), the big rip attractor relation
\begin{equation}
\label{atr}
\alpha=-3\phi + \alpha_0.
\end{equation}

To prove that the big rip solution (\ref{atr}) is an attractor, we will make the change of variables \cite{kie},
\begin{eqnarray}
\label{u}
u(\alpha,\phi)&\equiv &\frac{3}{10}e^{3\alpha-\phi}\left[\cos (\alpha + 3\phi)+\frac{1}{3}\sin (\alpha + 3\phi)\right],\nonumber\\
v(\alpha,\phi)&\equiv &\frac{3}{10}e^{3\alpha-\phi}\left[\sin (\alpha + 3\phi)-\frac{1}{3}\cos (\alpha + 3\phi)\right]
\end{eqnarray}
yielding the hamiltonian
\begin{equation}
\label{hamiltonianuv}
{\it{H}}=N\left(-\frac{1}{2}\pi^{2}_{u}-\frac{1}{2}\pi^{2}_{v}+1\right)e^{3\alpha-2\phi}.
\end{equation}
Its Hamilton-Jacobi equation reads
\begin{equation}
\left(\frac{\partial S_{0}}{\partial u}\right)^{2}+\left(\frac{\partial S_{0}}{\partial v}\right)^{2}=1,
\end{equation}
with solution
\begin{equation}
\label{acaoclassica}
S_{0k}=ku-\sqrt{\left(1-k^{2}\right)} v,
\end{equation}
and $|k|\leq 1$ is a constant of integration.

From the equations
\begin{eqnarray}
\label{clagui}
&&\dot{u}=-N\exp (3\alpha-2\phi)\pi_{u}=-N\exp (3\alpha-2\phi)\frac{\partial S_{0k}}{\partial u}=-N\exp (3\alpha-2\phi)k,\\
&&\dot{v}=-N\exp (3\alpha-2\phi)\pi_{v}=-N\exp (3\alpha-2\phi)\frac{\partial S_{0k}}{\partial v}=N\exp (3\alpha-2\phi)\sqrt{1-k^2},
\end{eqnarray}
and the definitions (\ref{u}), one obtains the first order equations
\begin{eqnarray}
\dot{\alpha}&=&e^{-\phi}\left[\sqrt{1-k^{2}}\sin\left(3\phi+\alpha\right)-k\cos\left(3\phi+\alpha\right)\right],\nonumber\\
\dot{\phi}&=&e^{-\phi}\left[k \sin\left(3\phi+\alpha\right)+\sqrt{1-k^{2}}\cos\left(3\phi+\alpha\right)\right].
	\label{fiponto}
\end{eqnarray}
One can easily verify that these equations are first integrals of the system (\ref{eqfriedmann}) and constraint
(\ref{hubble}), and are equivalent to them (in both cases we have three independent constants of integration).

From the Hamilton-Jacobi theory \cite{lanczos}, one obtains that 
\begin{equation}
\label{hj}
\frac{\partial S_{0k}}{\partial k}=\beta,
\end{equation}
where $\beta$ is a constant. Using again definitions (\ref{u}) in Eq.(\ref{hj}), one obtains
\begin{equation}
\label{intj}
\frac{3}{10}e^{3\alpha-\phi}\left[\frac{\sqrt{1-k^{2}}+3k}{3\sqrt{1-k^{2}}} \sin\left(3\phi+\alpha\right)+\frac{3\sqrt{1-k^{2}}-k}{3\sqrt{1-k^{2}}}\cos\left(3\phi+\alpha\right)\right]=\beta.
\end{equation}

Equation (\ref{intj}) is the implicit integral of the equation coming from Eq.(\ref{fiponto}),
\begin{equation}
\label{dj}
\frac{d\alpha}{d\phi}=\frac{\sqrt{1-k^{2}}\tan\left(3\phi+\alpha\right)-k}
{k\tan\left(3\phi+\alpha\right)+\sqrt{1-k^{2}}},
\end{equation}
as one can easily verify.

The big rip solution $\alpha + 3\phi=\alpha_0$ corresponds to $\beta = 0$, as one can check from
Eqs.(\ref{intj}) and (\ref{dj}). In this case, $\alpha_0$ and $k$ are related through
\begin{equation}
\label{aj}
\tan(\alpha_0)=\frac{k-3\sqrt{1-k^{2}}}{3k+\sqrt{1-k^{2}}},
\end{equation}
with solution $\alpha_0 = \alpha_{0k} + n\pi$ ($n$ an integer), for each $k$.

The field plot solutions of Eqs.(\ref{fiponto}) for the cases $k=0$ and $k=1/2$ are shown
in figures 1 and 2. Note the attractors at $\alpha + 3\phi=\alpha_0$ when $\phi\rightarrow-\infty, \alpha\rightarrow\infty$. 
The bounces at small values of $\alpha$ should not be considered because they are beyond the validity of the
model: in such regions, other matter components should be important while the scalar field should be irrelevant.

\begin{figure}
\includegraphics[width=8cm]{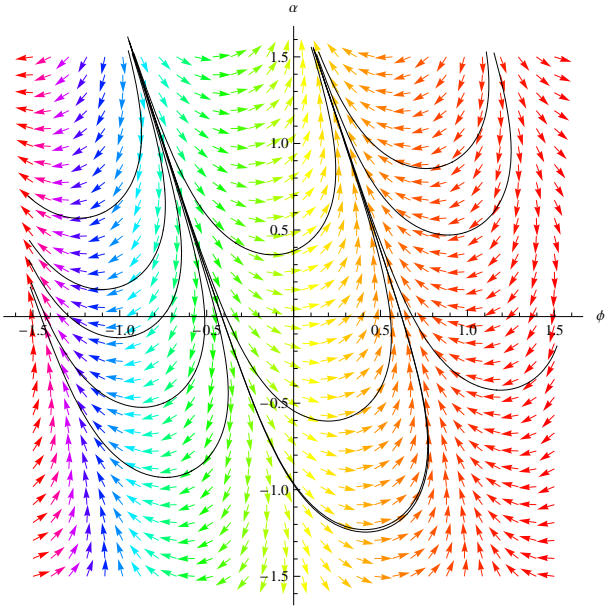}
\caption{Field plot figure showing $\alpha$ against $\phi$ for the classical case with $k=0$. Note the attractor behavior of the
curves $\alpha + 3\phi=\alpha_0$.}
\end{figure} 

\begin{figure}
\includegraphics[width=8cm]{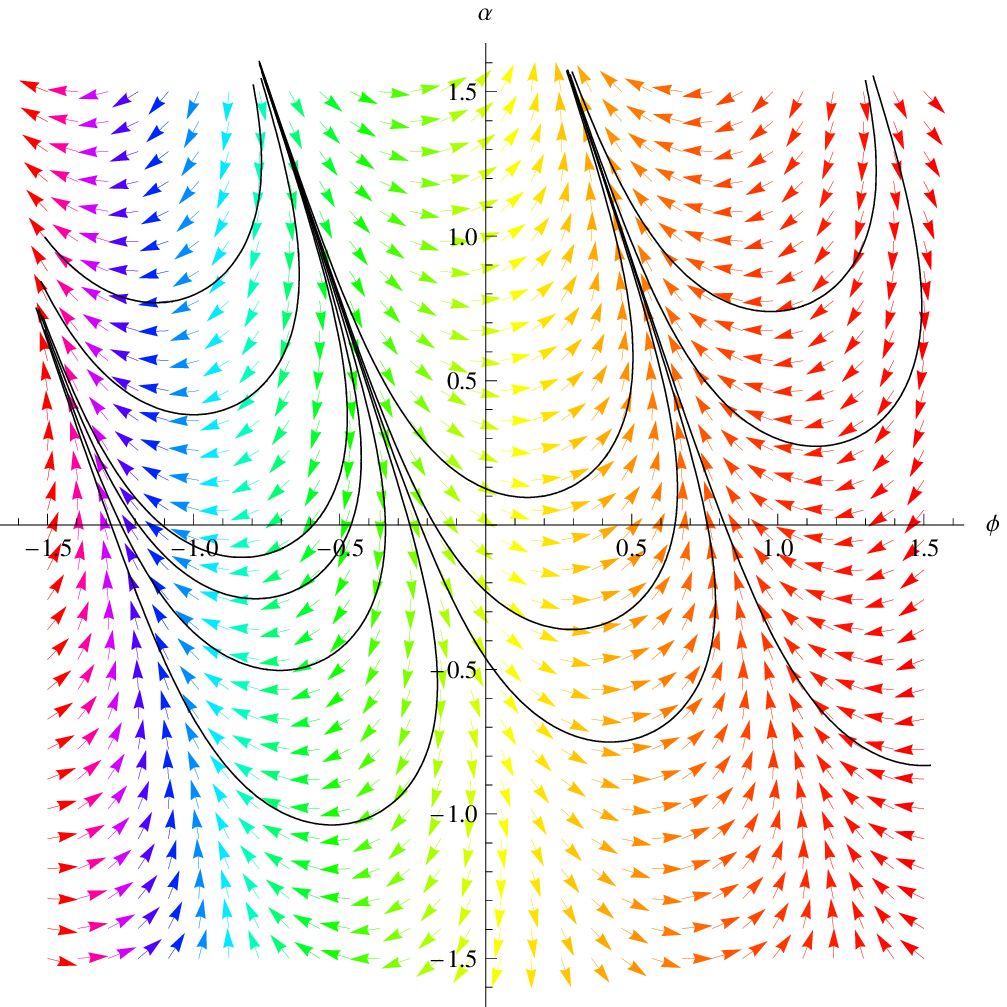}
\caption{Field plot figure showing $\alpha$ against $\phi$ for the classical case with $k=0.5$. Note again the attractor behavior of the curves $\alpha + 3\phi=\alpha_0$.}
\end{figure} 

One can understand analitically
their attractor behaviour in the folowing way: make the substitution 
$3\phi+\alpha \rightarrow \alpha_0 +\epsilon$ in Eq.(\ref{dj}) with $\alpha_0$ given in Eq.(\ref{aj}).
The result, up to first order in $\epsilon$, is 
\begin{equation}
\label{djeps}
\frac{d\alpha}{d\phi}=-3+10\epsilon+O(\epsilon^2),
\end{equation}
which means that the curves in the neighbhorhood of $\alpha + 3\phi=\alpha_0$, above and below, have inclination
in its direction. Hence they are attractors. 

Note that result (\ref{djeps}) is independent of $k$, the first term in the series is
linear and positive, and that
$\alpha_0$ given in Eq.(\ref{aj}) is the unique solution of $d\alpha/d\phi = -3$ (up to additions of $n\pi$).
These properties are not present in the quantum case, as we will see in the next section.

\section{Minisuperspace Bohm-de Broglie Quantization}

Dirac quantization of constrained systems imposes that the operator version of Hamiltonian (\ref{hamiltonian}) should
annihilate the wave function $\psi(a,\phi)$, yielding the Wheeler-Dewitt equation
\begin{equation}
\label{miniwheeler}
\left[\frac{\hbar^{2}}{2}\frac{\partial^{2}}{\partial\alpha^{2}} +\frac{\hbar^{2}}{2}\frac{\partial^{2}}{\partial \phi^{2}}+V_{0}e^{6\alpha-\lambda\sqrt{6}\phi}\right]\psi\left(\alpha,\phi\right)=0,
\end{equation}
which, in the variables $u$ and $v$, read 
\begin{equation}
	\hbar^{2}\left(\frac{\partial^{2}\psi}{\partial u^{2}}+\frac{\partial^{2}\psi}{\partial v^{2}}\right)+\psi=0.
\end{equation}

Using the Hamilton-Jacobi solution (\ref{acaoclassica}), one can construct the exact solution,
\begin{equation}
\label{solucao}
\psi\left(u,v\right)=C_{1}e^{\frac{i}{\hbar}\left[zu-\sqrt{\left(1-z^{2}\right)}v\right]}+C_{2}e^{-\frac{i}{\hbar}\left[zu-\sqrt{\left(1-z^{2}\right)}v\right]}.
\end{equation} 
Note that this solution is valid even for $|z|>1$.

One can construct gausssian superpositions of the above solution, 

\begin{equation}
\label{psigaussiano}
	\psi\left(u,v\right)=\int^{\infty}_{-\infty} dzA\left(z\right)\left\{C_{1}e^{\frac{i}{\hbar}\left[zu-\sqrt{\left(1-z^{2}\right)}v\right]}+C_{2}e^{-\frac{i}{\hbar}\left[zu-\sqrt{\left(1-z^{2}\right)}v\right]}\right\},
\end{equation}

where $A\left(z\right)$ is a Gaussian with width $\sigma$ centered around $\bar{z}$,

\begin{equation}
A\left(z\right)=\frac{\hbar}{\sigma\sqrt{2\pi}}e^{-\frac{\left(z-\bar{z}\right)^{2}}{2\sigma^{2}}\hbar^{2}}.
\end{equation}

Assuming that $\sigma$ is very small and setting $C_2=0$, one can perform the integration after an expansion around $z-\bar{z}$ yielding \cite{kie}  
\begin{equation}
\label{funcionaanalitico}
\psi=C_{1}\hbar^{2}\sqrt{\frac{1}{1-i\sigma^{2}\hbar S^{''}_{0}}}\exp\left[\frac{iS_{0}}{\hbar}-\frac{S^{'2}_{0}}{2\left(\sigma^{-2}-i\hbar S^{''}_{0}\right)}\right],
\end{equation}
where $S_0(z)=zu-\sqrt{\left(1-z^{2}\right)}v$, and the primes are derivatives with respect to $z$. All functions
of $z$ are evaluated at $\bar{z}$.
Writting $\psi$ in polar form, $\psi=R e^{i\frac{S}{\hbar}}$, one obtains
\begin{eqnarray}
\label{raio}
&&R=\frac{C_{1}\hbar^{2}}{\left(1+\sigma^{4}\hbar^{2}S^{''2}_{0}\right)^{1/4}}\exp\left[-\frac{S^{'2}_{0}\sigma^{-2}}{2\left(\sigma^{-4}+\hbar^{2}S^{''2}_{0}\right)}\right],\nonumber \\
&&S=S_{0}-\frac{\hbar^{2} S^{'2}_{0}S^{''}_{0}}{2\left(\sigma^{-4}+\hbar^{2}S^{''2}_{0}\right)}+\frac{\hbar}{2}\arctan\left(\sigma^{2}\hbar S^{''}_{0}\right).
\label{faseS}
\end{eqnarray}

The explicit expression for $S$ reads,

\begin{equation}
S\left(u,v\right)=ku-\sqrt{(1-k^2)}v-\frac{(1-k^2)^{3/2}v\sigma^4\left(u+\frac{kv}{\sqrt(1-k^2)}\right)^{2}}{2[(1-k^2)^3+\sigma^{4}v^{2}]}+\frac{1}{2}\arctan\left[\frac{\sigma^{2} v}{(1-k^2)^{3/2}}\right],
\label{faseultima}
\end{equation}
where were we have set $\hbar=1$ and $\bar{z}\equiv k$. We will restrict ourselves to the case $|k|<1$ in order to avoid divergences
of the wave function.

Assuming the Bohm-de Broglie interpretation of quantum cosmology \cite{nelson}, we will use the Bohmian guidance relations
\begin{eqnarray}
\label{quagui}
\dot{u}&=&-N\exp (3\alpha-2\phi)\pi_{u}=-N\exp (3\alpha-2\phi)\frac{\partial S}{\partial u},\nonumber \\
\dot{v}&=&-N\exp (3\alpha-2\phi)\pi_{v}=-N\exp (3\alpha-2\phi)\frac{\partial S}{\partial v},
\end{eqnarray}
which have the same form as their classical counterpart (\ref{clagui}), except for the fact that the Hamilton-Jacobi function
which appears in (\ref{quagui}) is the quantum one given in Eq.(\ref{faseultima}), 
not the classical one given in Eq.(\ref{acaoclassica}), which implies the quantum effects.
Note that $S$ written in Eq.(\ref{faseultima}) reduces to $S_{0k}$ given in Eq.(\ref{acaoclassica}) when $\sigma\rightarrow 0$,
as expected, yielding the classical limit.

Going back to the original variables $\alpha,\phi$ through Eqs.(\ref{u}), Eq.(\ref{quagui}) yields,
\begin{eqnarray}
\dot{\alpha}&=&-e^{-\phi}\left[\frac{\partial S}{\partial v}\sin\left(3\phi+\alpha\right)+\frac{\partial S}{\partial u}\cos\left(3\phi+\alpha\right)\right],\nonumber\\
\dot{\phi}&=&e^{-\phi}\left[\frac{\partial S}{\partial u}\sin\left(3\phi+\alpha\right)-\frac{\partial S}{\partial v}\cos\left(3\phi+\alpha\right)\right],
	\label{qfiponto}
\end{eqnarray}
where it is understood that Eqs.(\ref{u}) must be used in the partial derivatives of $S$.
 
In the relevant region $\phi\rightarrow-\infty$, $\alpha\rightarrow\infty$, which corresponds to
$u\rightarrow\infty$, $v\rightarrow\infty$, Eq.(\ref{faseultima}) reduces to
\begin{equation}
S\left(u,v\right)=ku-\sqrt{(1-k^2)}v-\frac{(1-k^2)^{3/2}\left(u+\frac{kv}{\sqrt(1-k^2)}\right)^{2}}{2v} + \frac{\pi}{4}.
\label{limfaseultima}
\end{equation}
Note that it is independent of $\sigma$. Hence, we obtain,
\begin{eqnarray}
\frac{\partial S}{\partial u} &=&-\frac{\tan(\alpha+3\phi)+3}{3\tan(\alpha+3\phi)-1}(1-k^2)^{3/2}+k^3,\nonumber\\
\frac{\partial S}{\partial v} &=&-\frac{\sqrt{1-k^2}}{2}\left[2+k^2-\left(\frac{\tan(\alpha+3\phi)+3}{3\tan(\alpha+3\phi)-1}\right)^2(1-k^2)\right].
\label{qds}
\end{eqnarray}
These terms can now be substituted in Eqs.(\ref{qfiponto}) in order to obtain the quantum trajectories.

The behaviors of the quantum trajectories around the big rip solutions $\alpha + 3\phi=\alpha_0$ 
can be understood analitically as follows: from Eqs.(\ref{qfiponto},\ref{qds}) one obtains
\begin{equation}
\label{qdj}
\frac{d\alpha}{d\phi}= -\frac{-(17+10k^2)\sqrt{1-k^2} x^3+6[3 k^3+(k^2+2)\sqrt{1-k^2}] x^2+3[-4 k^3+(2k^2-3)\sqrt{1-k^2}] x+2 k^3-6(k^2+1)\sqrt{1-k^2}}{6[-(1-k^2)^{3/2}+3 k^3]x^3+[-12 k^3+(26k^2+1)\sqrt{1-k^2}] x^2+2[k^3-3(k^2+2)\sqrt{1-k^2}] x-(7-10k^2)\sqrt{1-k^2}},
\end{equation}
where $x=\tan(\alpha+3\phi)$.
The big rip solution $\alpha + 3\phi=\alpha_0$ can be obtained from the solutions $d\alpha/d\phi = -3$ of Eq.(\ref{qdj}),
as in the classical case. The difference in the quantum case is that now we have two roots given by
\begin{eqnarray}
\label{qaj}
\tan(\alpha_{01})&=&\frac{(36k^2+3)\sqrt{1-k^2}  +2k(10-k^2)}{(28k^2-1)\sqrt{1-k^2} +54 k^3},\nonumber \\
\tan(\alpha_{02})&=&-\frac{(54k^2-3) \sqrt{1-k^2}  +2k(5-14 k^2)}{(28 k^2-1)\sqrt{1-k^2} +54 k^3},
\end{eqnarray}
up to additions of $n\pi$.

Making again the substitution 
$3\phi+\alpha \rightarrow \alpha_{0i} +\epsilon$ in Eq.(\ref{qdj}), with $\alpha_{0i}$ given in Eq.(\ref{qaj}),
the result, up to leading order in $\epsilon$, is 
\begin{equation}
\label{q1djeps}
\frac{d\alpha}{d\phi}=-3+f(k)\epsilon+O(\epsilon^2),
\end{equation}
for $\alpha_{01}$, and
\begin{equation}
\label{q2djeps}
\frac{d\alpha}{d\phi}=-3+g(k)\epsilon^2+O(\epsilon^3),
\end{equation}
for $\alpha_{02}$, where
\begin{equation}
\label{fk}
f(k)=\frac{90k^2}{3k^2-2},
\end{equation}
and
\begin{equation}
\label{gk}
g(k)=-\frac{15k}{\sqrt{1-k^2}}.
\end{equation}

When $k=0$, the roots coincide and read, $\tan(\alpha_0)=-3$.
Both $f(k)$ and $g(k)$ are zero and one has to go to the third order, yielding
\begin{equation}
\frac{d\alpha}{d\phi}=-3-5x^3+O(x^4).
\end{equation}
Hence, the curves in the neighbhorhood of $\alpha + 3\phi=\alpha_0$, above and below it, 
have inclination contrary to its direction: it is a repellor and the big rip is avoided.

For all other $k$, the situation is more involved. In the case of the big rip curves corresponding to the first root $\alpha_{01}$, 
one can see from Eq.(\ref{q1djeps}) that the lines $\alpha + 3\phi=\alpha_{01} + n\pi$
are repellors or attractors, depending on the sign of $f(k)$. For $|k|<(2/3)^{1/2}$, $f(k)<0$ which means that the curves are repellors,
while for $(2/3)^{1/2}<|k|<1$, $f(k)>0$ and the curves are attractors.

For the second root $\alpha_{02}$, Eq.(\ref{q2djeps}) indicates that the curves $\alpha + 3\phi=\alpha_{02} + n\pi$ 
are "saddle" lines, which work as attractors from one side and as repellors from the other side. 
For $k>0$, $f(k)<0$ which means that the curves are repellors from above and attractors from below,
while for $k<0$, $f(k)>0$ and the curves are attractors from above and repellors from below.

From these considerations we conclude that, for $|k|<(2/3)^{1/2}$, half of the initial conditions giving rise to
expanding solutions go to the big rip if they are to the left (right) of $\alpha_{02}$, and avoid the big rip
if they are to the right (left) of $\alpha_{02}$ if $k>0$ ($k<0$). When $(2/3)^{1/2}<|k|<1$, all expanding solutions 
go to the big rip, even after being repelled around $\alpha_{02}$.

These behaviours can be seen numerically in Figs. 3, 4, 5, 6, corresponding to the cases $k=0$, $k=1/2$, $k=-1/2$ and $k=9/10$, 
respectively, which are representative of all possible cases described above.

\begin{figure}
\includegraphics[width=8cm]{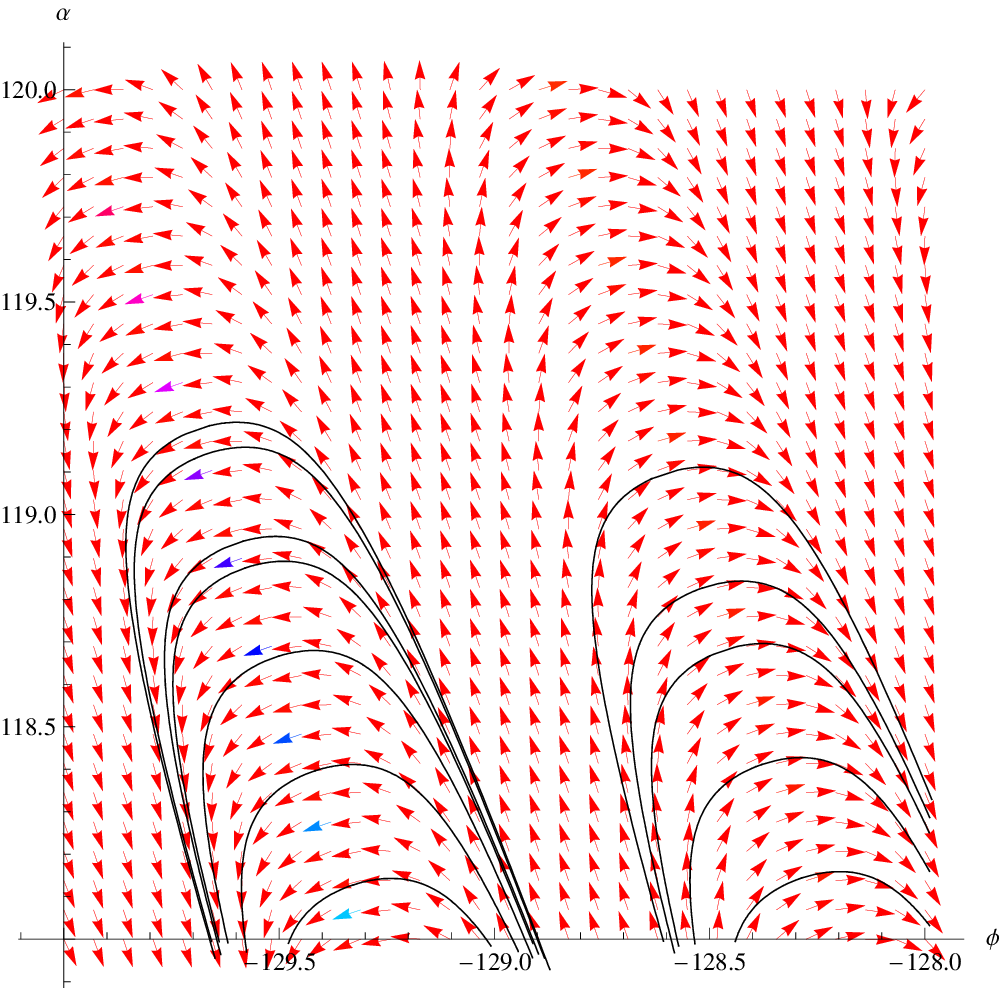}
\caption{Field plot figure showing $\alpha$ against $\phi$ for the quantum case with $k=0$. Note the repellor behavior of the
curves $\alpha + 3\phi=\alpha_0$.}
\end{figure}

\begin{figure}
\includegraphics[width=8cm]{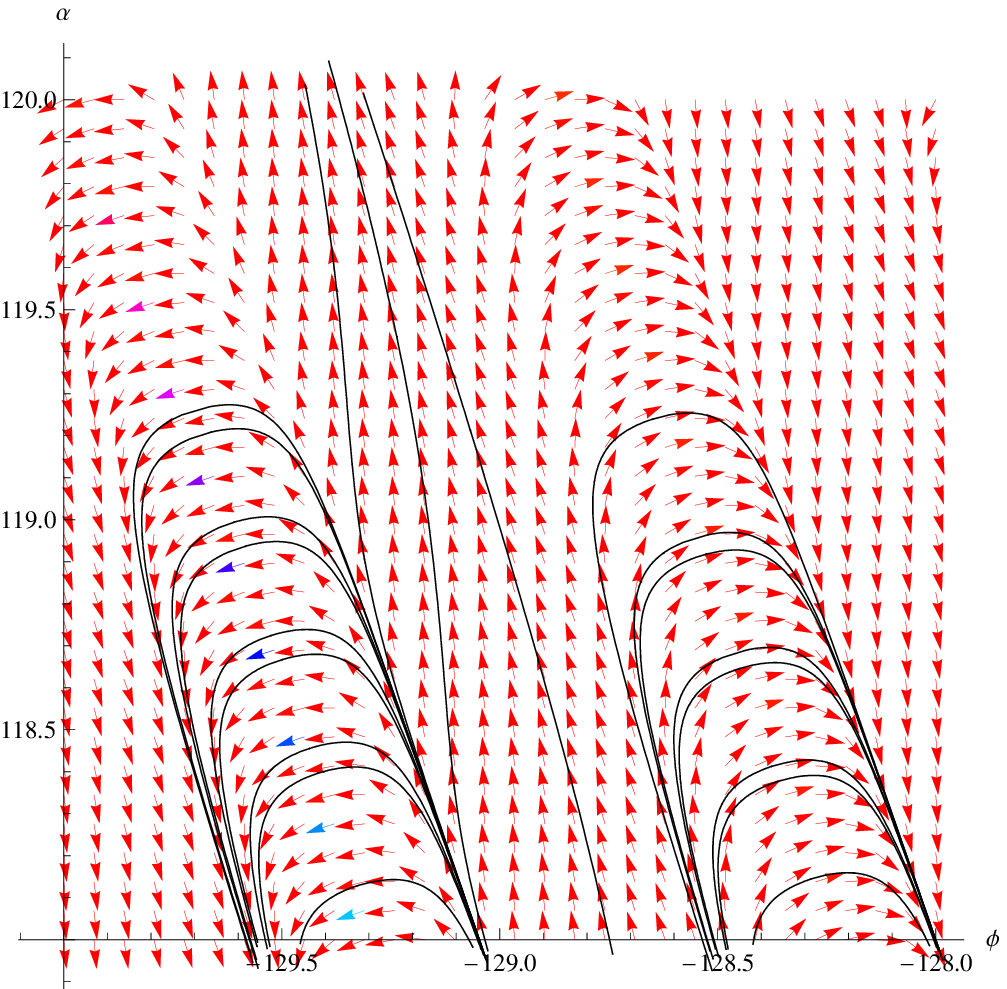}
\caption{Field plot figure showing $\alpha$ against $\phi$ for the quantum case with $k=0.5$. Note the saddle curves
which works as attractors from above and repellors from below, as predicted from our analitic discussion.}
\end{figure}

\begin{figure}
\includegraphics[width=8cm]{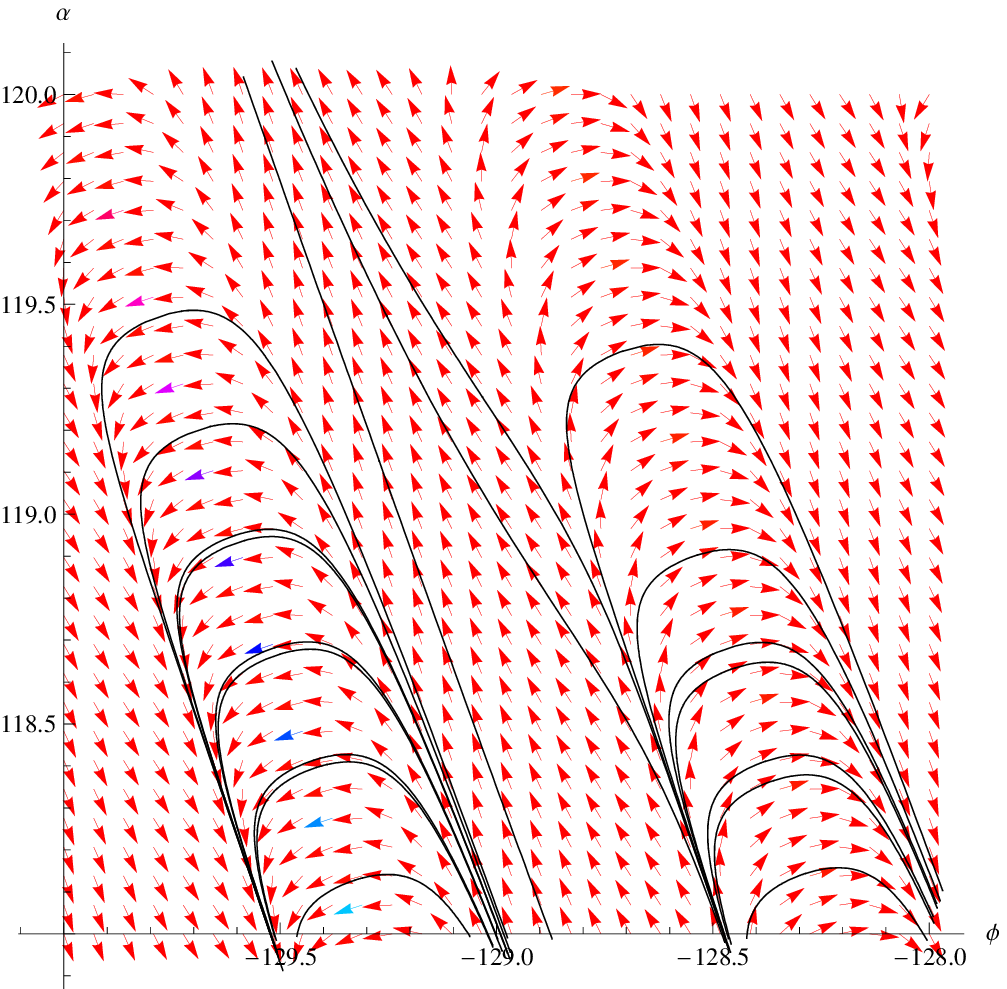}
\caption{Field plot figure showing $\alpha$ against $\phi$ for the quantum case with $k=-0.5$. Note the saddle curves
which works as attractors from below and repellors from above, as predicted from our analitic discussion.}
\end{figure}

\begin{figure}
\includegraphics[width=8cm]{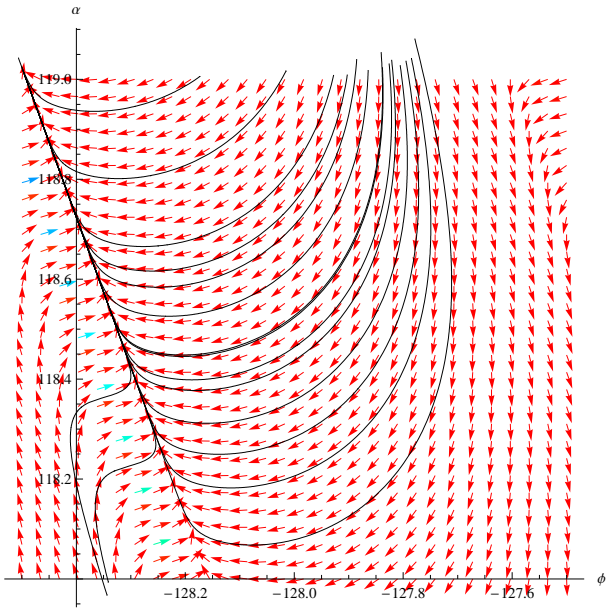}
\caption{Field plot figure showing $\alpha$ against $\phi$ for the quantum case with $k=0.9$. One can see the attractor behaviour
of the big rip, as predicted from our analitic discussion.}
\end{figure}

\section{Conclusion}

We have shown in this paper that minisuperspace quantization of homogeneous and isotropic geometries
with phantom scalar fields, when examined in the light of the Bohm-de Broglie interpretation of quantum mechanics,
does not eliminate, in general, the classical big rip singularity. When examining gaussian superpositions of exact solutions
of the Wheeler-DeWitt equation centered at different values of the Hamilton-Jacobi separation constant $|\bar{z}|\equiv |k|<1$, 
one arrives at three possibilities: the big rip is completely eliminated when $k=0$, it is a future attractor of all quantum expanding solutions
when $(2/3)^{1/2}<|k|<1$, and it can be either eliminated or a future attractor, depending on the initial conditions of each Bohmian trajectory,
when $|k|<(2/3)^{1/2}$. As in the classical regime the big rip is a future attractor for all values of the separation constant $k$,
one concludes that, using the Bohm-de Broglie interpretation, the big rip singularity problem is alleviated but not solved by quantization.
 
However, contrary to this conclusion, it is argued in Ref.\cite{kie}, using a different interpretation of the wave function, 
that the big rip singularity is completely eliminated ("smoothed out") through quantization, and this result is independent 
of the separation constant $k$. The reason for these different conclusions relies on the fact that all information 
one can get from the interpretation used in Ref.\cite{kie} comes
from the amplitude of the wave function, whose qualitative behavior does not depend 
on $k$, while the results obtained in the present paper, where the Bohm-de Broglie interpretation of quantum mechanics
was used, there is crucial information also coming from the phase of the wave function (its gradient), whose qualitative behavior does  
depend on $k$. We are then faced with the very peculiar situation where different interpretations of 
the same quantum state of a system are predicting different physical facts, instead of just giving different 
descriptions of the same observable facts: indeed, there is nothing more observable than the fate of the whole Universe. 
Of course, even if the toy model analyzed here has something to do with the real Universe, one should wait some tens
of billion years to decide which conclusion is correct. 
We have explained in the Introduction why we think the results of Ref.\cite{kie} are not conclusive, and how
the notion of Bohmian trajectories coming from the Bohm-de Broglie interpretation of quantum mechanics leads
to well posed questions and answers concerning this subject. Note, however, that the very notion of Bohmian
trajectories can be questioned with the argument that they are just artifacts, with no physical meaning.
In fact, many objections against this notion have been presented \cite{scul}, which were however properly answered
in Ref.\cite{hil}. Anyway, deciding between these two results seems to be rather premature at the moment.

Let us conclude with a last remark. We have seen that in quantum cosmology one may arrive at the
peculiar situation where different interpretations of the same quantum state can lead to different
physical facts, although testing the alternatives is completely out of question in this case. 
One interesting perspective for future work should be to find analog models 
in the laboratory based on this (and perhaps others) quantum cosmological model which
present analog ambiguities in order to decide between interpretations. If it indeed turns to be
possible in the future, it will be a result of ultimate importance for quantum mechanics,
and for our whole physical picture of reality.

\section*{Acknowledgements} We would like to thank CNPq of Brazil for
financial support, and Claus Kiefer for his comments and suggestions.

\end{document}